\documentclass{article}
\usepackage[utf8]{inputenc}
\usepackage[super,sort&compress]{natbib}
\usepackage{authblk}
\usepackage{setspace}
\usepackage{amsmath}
\usepackage{amsfonts}
\usepackage{braket}
\usepackage{comment}
\usepackage[normalem]{ulem}
\usepackage{graphicx}
\usepackage{physics}
\usepackage{xcolor}
\usepackage{float}
\usepackage{multirow}
\usepackage{adjustbox}
\usepackage{hyperref}
\usepackage{enumitem,kantlipsum}
\usepackage{upgreek}
\usepackage{soul}
\usepackage{geometry}
\usepackage{gensymb}
\usepackage[normalem]{ulem}
\title{Retrieving intrinsic polarization anisotropies of nanostructures using differential Mueller matrix polarimetry}
\author[1]{Jeeban Kumar Nayak\thanks{jeeban.nayak@epfl.ch}}
\author[1]{Ebru Buhara}
\author[1]{Olivier J.~F.~Martin\thanks{olivier.martin@epfl.ch}}
\affil[1]{Nanophotonics and Metrology Laboratory (NAM), Swiss Federal Institute of Technology Lausanne (EPFL), Lausanne 1015, Switzerland}
\date{}
\begin{document}
\maketitle
\section*{Abstract}
Accurate characterization of polarization-dependent light–matter interactions in nanostructured systems is paramount for the development of chiral metasurfaces. It is also often challenging, because multiple anisotropic mechanisms -- such as linear and circular diattenuation, birefringence, and depolarization can coexist and couple with one another. Conventional ellipsometric and chiro-optical techniques typically assume isolated polarization effects and can therefore yield inaccurate estimates of the intrinsic polarization parameters. Here, we demonstrate that Mueller matrix polarimetry combined with a differential Mueller matrix decomposition provides a robust framework for retrieving the intrinsic polarization response of complex nanophotonic systems. Using plasmonic gammadion arrays and media with multiple polarization anisotropies as multi-modal chiral platforms, we show that simultaneous linear and circular anisotropies produce coupled signatures in the Mueller matrix, leading to significant artifacts in conventional polarization observables. Through analytical modeling and experimental measurements, we quantify these artifacts and demonstrate that a differential decomposition accurately decouples and retrieves the underlying polarization parameters. The presented approach also successfully probes the polarization anisotropic effects in inhomogeneous media enabling a clear discrimination between the intrinsic chiral optical response and geometric phase effects arising from spin–orbit interaction of light in momentum-resolved scattering. These results establish differential Mueller matrix polarimetry as a powerful tool for rigorous characterization of polarization phenomena in nanostructured photonic systems and polarization-engineered metasurfaces.
\section*{Introduction}
Among the various degrees of freedom of light, polarization has emerged as a powerful resource in nanophotonics for probing light–matter interactions and engineering advanced optical functionalities \cite{hu2020allpolmani2,li2024opticalpolmani1,martinez2018polarimetrynano,novotny2011antennas,yu2014flatopticspolmetareview,shaltout2019spatiotemporalmetareview2,chen2021principles}. Tailoring the polarization response of nanostructures has enabled the observation of a wide range of fundamental physical phenomena, including optical chirality \cite{electromagneticchiralityreview1,valev2013chirality1}, anisotropic scattering \cite{hu2020allpolmani2,li2024opticalpolmani1}, spin–orbit interactions \cite{bliokh2015spinsoi1}, structured light generation \cite{he2022towardsstructuredlight}, and polarization-dependent resonances \cite{doeleman2018experimentalpolbic}, to name a few. These capabilities have further opened pathways toward numerous practical applications, such as precise manipulation of light at the nanoscale \cite{hu2020allpolmani2,li2024opticalpolmani1,bandaru2022microstructure}, ultrasensitive molecular detection \cite{im2024perspectivessensingreview1,liu2023detectionbiomreview3sensing}, chiral lasing \cite{peng2016chirallasing}, polarization-multiplexed optical communication \cite{chvojka2020visiblepolarizationmultiplexing}, data storage \cite{lamon2024neuromorphicdata}, and several polarization-engineered nanophotonic devices \cite{shen2023reviewpoldevices1}. Polarization-resolved measurements provide direct access to these optical phenomena. Consequently, the rapidly growing research efforts in nanophotonics demand robust and comprehensive polarimetric platforms capable of accurately probing the polarization responses of nanostructured systems \cite{li2024opticalneedpol,shao2022analyzingneedpol2,ray2017polarizationmmnano1}.
\par
Polarization ellipsometry has been extensively employed to characterize polarization effects in nanostructures by measuring changes in the orientation and ellipticity of the scattered or transmitted polarization state \cite{kawabata2014fundamentalsellipsometry,losurdo2009spectroscopicellipsometry2,schubert2004infraredellipsometry3}. Following a similar methodology, conventional chiro-optical measurements quantify chirality through circular diattenuation (CD) and circular birefringence (CB) \cite{hentschel2017chiralchiro-opticalnano,schaferling2012tailoringchiro-opticalnano2}. However, most conventional chiro-optical and ellipsometric techniques implicitly assume that a single anisotropic mechanism dominates the polarization response. In nanostructured systems where multiple anisotropies coexist and interact, such assumptions can lead to inaccurate interpretation of the measured polarization signatures \cite{miao2025opticalchiro-opticalmm,ugras2023canchiralartifat2,lightner2023understandingchiralartifact1}. In complex nanostructures, linear anisotropy, circular anisotropy, and depolarization may occur simultaneously and couple with one another. Under such conditions, the utilization of a standard polarimetric approach fails to probe intrinsic polarization parameters \cite{yao2022extractingchiro,lightner2023understandingchiralartifact1,ugras2023canchiralartifat2}, quantities that are critically important for applications such as chiral sensing and polarization-selective nanophotonic devices \cite{im2024perspectivessensingreview1}.
\par
Mueller matrix (MM) polarimetry provides a more complete framework in this regard, as it captures the full polarization response of a system through a $4 \times 4$ matrix representation \cite{goldstein2017polarized,gil2022polarizedpolbook1,gupta2015waveoptics}. Each element of the MM encodes specific polarization interactions, enabling simultaneous access to the fundamental polarization parameters : diattenuation, retardance, depolarization, as well as their mutual coupling. Simultaneous presence of these basic polarization interactions may yield non-trivial polarization manifestations \cite{goldstein2017polarized}. For instance, in a recent work, coupling between linear diattenuation and linear birefringence effects led to circular polarization-dependent absorption \cite{parrish2025differentialldlbcoupling}, which is conventionally treated as optical activity. Accordingly, the concurrent occurrence of multiple polarization effects often manifests in an interrelated manner in the MM elements. In such scenarios, interpretation of raw MM data alone is insufficient; systematic decomposition is required to isolate the intrinsic polarization parameters.
\par
In the recent past, several MM decomposition techniques have been developed, among which polar decomposition and differential decomposition are widely explored \cite{lu1996interpretationpolardecomp1,arteaga2009pseudopolar}. The polar decomposition expresses the MM as a sequential product of basis matrices representing diattenuation, retardance, and depolarization \cite{goldstein2017polarized,gupta2015waveoptics}. However, this factorization is not unique, and the extracted parameters may depend on the assumed ordering of the constituent matrices, potentially introducing ambiguities in systems with multiple anisotropies \cite{kumar2024muellerpolarvsdiff,kumar2012comparativepolarvsdiff}. In contrast, differential decomposition \cite{azzam1978propagationdiffdecomp1,ossikovski2011differential} treats all polarization effects as simultaneously occurring processes and represents them within a single differential matrix formalism. This approach inherently accounts for the mutual coupling between anisotropic contributions and therefore might be well suited for complex nanostructures exhibiting multiple concurrent polarization mechanisms \cite{yu2014flatmultipolnano}.
\par
In this work, we combine spectroscopic Mueller matrix polarimetry with differential Mueller matrix decomposition to rigorously retrieve intrinsic polarization parameters in nanostructured systems exhibiting multiple coupled anisotropies.
Specifically, we (i) quantify the artifacts that arise in conventional ellipsometric and chiro-optical observables when multiple anisotropies coexist, (ii) demonstrate that differential Mueller matrix decomposition accurately retrieves intrinsic polarization parameters without sequential assumptions, and (iii) apply this framework to experimentally measure the linear and circular polarization effects associated with plasmonic gammadion arrays. Furthermore, we demonstrate the effectiveness of MM polarimetry in capturing inhomogeneous polarization scattering typically observed in nanostructured systems, providing deeper physical insight into the underlying light–matter interactions and their manifestation in measurable polarization signatures.
\section*{Theory}
MM polarimetry is a comprehensive framework for describing polarization-dependent light–matter interactions. The technique measures the $4\times4$ Mueller matrix $M$, which relates the input and output Stokes vectors through \cite{goldstein2017polarized,gupta2015waveoptics}
\begin{equation}
\mathbf{S_{\mathrm{out}} = M S_{\mathrm{in}}},
\end{equation}
The Stokes vector is defined as
\begin{equation}
\begin{pmatrix}
S_0 \\
S_1 \\
S_2 \\
S_3
\end{pmatrix}
=
\begin{pmatrix}
I_H + I_V \\
I_H - I_V \\
I_P - I_M \\
I_L - I_R
\end{pmatrix}.
\end{equation}
Here $I_H$ and $I_V$ denote horizontal and vertical linear polarization intensities, $I_P$ and $I_M$ correspond to $\pm 45^\circ$ linear polarizations, and $I_L$ and $I_R$ represent left- and right-circularly polarized intensities.
All kinds of polarization effects present in the medium are encoded in the MM. In a typical scenario, the elements in the first row quantify diattenuation, i.e., differential attenuation for orthogonal polarization states. Conversely, the elements in the first column describe polarizance, representing the ability of the medium to convert unpolarized light into polarized light. While these elements may appear similar for ideal polarizers, they encode different physical processes depending on whether the polarization selectivity arises from differential attenuation or polarization generation. This distinction can be further understood from the polarization-resolved intensities used to construct the individual MM elements, as illustrated in Fig.~\ref{fig1}(a). 
For instance, the total transmitted intensity under unpolarized illumination can be expressed as $HH+HV+VH+VV = UU$, while combinations such as $HH+HV-VH-VV=HU-VU$ capture differential transmittance between horizontal and vertical polarized excitation. Consequently, it can be seen that the elements in the first row quantify the difference in total transmitted intensity for input orthogonal linear or circular polarization states, whereas the elements in the first column represent the difference between orthogonally polarized out-put intensities under unpolarized illumination.  
The CD effect is described by $M_{14}$ and $M_{41}$, whereas linear diattenuation is associated with $M_{12}$ and $M_{13}$. Linear retardance, corresponding to phase accumulation between orthogonal linear polarization states, manifests primarily in $M_{24}$, $M_{42}$, $M_{34}$, and $M_{43}$, which govern linear–circular polarization conversion. The CB effect, accounting for the opposite phase accumulation between left-circularly polarized (LCP) and right-circularly polarized (RCP) light, manifests as the rotation of linearly polarized light and appears in the off-diagonal elements $M_{23}$ and $M_{32}$ \cite{goldstein2017polarized,gupta2015waveoptics}.
In the absence of depolarization, the diagonal elements are determined by the magnitude and orientation of these anisotropic effects; however, depolarization modifies both the diagonal and off-diagonal elements \cite{ossikovski2011differential}.
\par
Standard ellipsometric and chiro-optical measurements often rely on symmetry assumptions of simplified Mueller matrices. For instance, an ideal non-depolarizing medium exhibiting only circular anisotropy (CD and CB) can be represented as
\begin{equation}
\mathbf{M_c} =
\begin{pmatrix}
1 & 0 & 0 & CD \\
0 & \cos 2\theta & \sin 2\theta & 0 \\
0 & -\sin 2\theta & \cos 2\theta & 0 \\
CD & 0 & 0 & 1
\end{pmatrix},
\end{equation}
where $\theta$ denotes the optical rotation arising from circular birefringence.
For an incident $+45^\circ$ linearly polarized state,
\begin{equation}
\mathbf{S_{\mathrm{in}}} =
\begin{pmatrix}
1 \\ 0 \\ 1 \\ 0
\end{pmatrix},
\end{equation}
the output Stokes vector becomes
\begin{equation}
\mathbf{S_{\mathrm{out}} = M_c S_{\mathrm{in}}} =
\begin{pmatrix}
1 \\
\sin 2\theta \\
\cos 2\theta \\
CD
\end{pmatrix}.
\end{equation}
In this idealized case, the polarization orientation angle directly yields the CB parameter $\theta$, while the induced ellipticity determines CD \cite{yao2022extractingchiro}. However, this simplified extraction is valid only for non-depolarizing media exhibiting sole circular anisotropy. In realistic systems -- such as biological samples, molecular assemblies, or plasmonic nanostructures --multiple anisotropic mechanisms (linear diattenuation, linear birefringence, circular effects, and depolarization) coexist and couple. These interactions break the simplified symmetry of $M_c$, leading to crosstalks between Mueller matrix elements. Consequently, direct extraction of CD and CB from conventional ellipsometric observables can result in significant artifacts. Accurate quantification therefore requires to separate and quantify intrinsic individual polarization contributions \cite{ossikovski2011differential}.
\par
To decouple simultaneously occurring polarization effects, we employ the differential Mueller matrix formalism. In this approach, the local polarization properties of a medium are described by a differential matrix $m$, related to the spatial evolution of the Mueller matrix along the propagation direction $z$ as \cite{ossikovski2011differential}
\begin{equation}
\frac{d\mathbf{M}}{dz} = \mathbf{m.M}
\end{equation}
For a medium with uniform polarization properties along the optical path length $l$, integration yields
\begin{equation}
L = \ln \mathbf{M} = \textbf{m} l,
\end{equation}
where $\mathbf{L}$ is the logarithmic Mueller matrix.
The matrix $\mathbf{L}$ can be decomposed into Lorentz antisymmetric ($\mathbf{L_m}$) and Lorentz symmetric ($\mathbf{L_u}$) components \cite{ossikovski2011differential}:
\begin{equation}
\mathbf{L_m} = \mathbf{\frac{1}{2}(L - G L^{T} G)}, \quad
\mathbf{L_u} = \mathbf{\frac{1}{2}(L + G L^{T} G)},
\end{equation}
where $\mathbf{G} = \mathrm{diag}(1,-1,-1,-1)$ is the Minkowski metric tensor.

The antisymmetric component $\mathbf{L_m}$ contains the intrinsic (non-depolarizing) polarization anisotropies, while $\mathbf{L_u}$ accounts for depolarization effects and statistical uncertainties arising from inhomogeneities. For a non-depolarizing medium, $\mathbf{L_m}$ takes the general form \cite{ossikovski2011differential}
\begin{equation}
\mathbf{L_m} =
\begin{pmatrix}
0 & -LD & -LD' & CD \\
-LD & 0 & CB & LB' \\
-LD' & -CB & 0 & -LB \\
CD & -LB' & LB & 0
\end{pmatrix}.
\end{equation}
Here $LD$ and $LD'$ represent linear diattenuation along the horizontal/vertical and $\pm 45^\circ$ bases, respectively; $LB$ and $LB'$ denote the corresponding linear birefringence components; and $CD$ and $CB$ correspond to circular diattenuation and circular birefringence. 
Thus, the differential formalism enables direct extraction of intrinsic anisotropic parameters from the logarithmic Mueller matrix without assuming sequential ordering of effects. This framework is particularly advantageous for nanostructured systems where multiple polarization mechanisms coexist and interact simultaneously.
\section*{Results}
\begin{figure*}[htbp]
\centering
    \includegraphics[width=\linewidth]{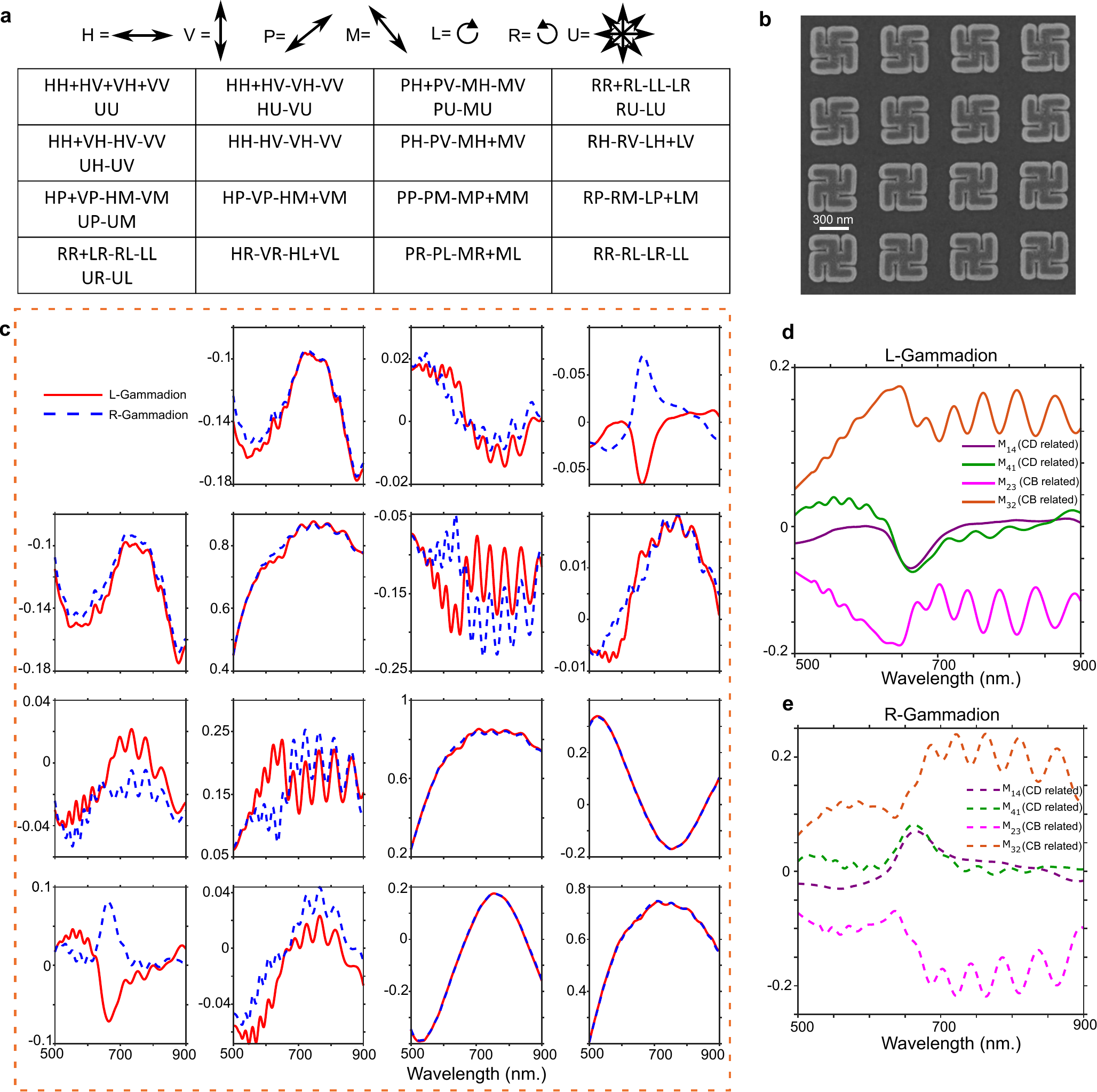}
 \caption{Spectral Mueller matrix characterization for left- and right-handed periodic gammadion arrays. (a) Measurement scheme used to reconstruct the Mueller matrix from 36 polarization-resolved projective measurements. The first letter denotes the incident polarization state and the second letter denotes the analyzed polarization state. The polarization states are defined as $I_H$ (horizontal), $I_V$ (vertical), $I_P$ $(+45^\circ)$, $I_M$ $(-45^\circ)$, $I_L$ left circular polarization (LCP), and $I_R$ right circular polarization (RCP). (b) Scanning electron microscopy images of both left and right handed gammadion arrays. (c) Normalized ($4\times4$) spectral Mueller matrices measured for the left-handed (L, solid red line) and right-handed (R, blue dashed line) gammadion arrays. (d,e) Spectral evolution of the circular anisotropy descriptor Mueller matrix elements. The elements $M_{14}$ and $M_{41}$ quantify circular diattenuation (CD), whereas $M_{23}$ and $M_{32}$ describe circular birefringence (CB), highlighting the chiro-optical response of the plasmonic gammadion arrays.}
\label{fig1}
\end{figure*}

To demonstrate the power of MM polarimetry to study chiral resonant metasurfaces, we employ spectral-domain MM polarimetry to investigate polarization-dependent light–matter interactions in periodic plasmonic gammadion arrays with well-defined structural handedness, exemplified in Fig.~\ref{fig1}(b). The geometrical parameters of the gammadion are designed to support strong chiral resonances in the visible spectral range (600 -- 700 nm).The gold gammadion arrays are fabricated with a periodicity of $800$ nm in both x and y directions, on a thin Glass substrate. The nanostructures are fabricated by means of electron beam lithography and subsequent etching technology reported elsewhere \cite{abasahl2021fabrication}. The side length of each gammadion is 240 nm. The observed chiro-optical response originates from the excitation and interference of higher-order electric and magnetic multipole modes. Multipolar decomposition of the scattered fields reveals dominant contributions from electric octupole and magnetic quadrupole moments, whose coupling produces different responses for left- (LCP) and right-circularly polarized (RCP) excitations, resulting in measurable circular diattenuation and birefringence. A detailed discussion on the excitation of multipolar electromagnetic modes and their role in optical chirality can be found in the literature \cite{electromagneticchiralityreview1}.
\par
The experimental polarimetry configuration consists of a custom polarization-resolved microscopic setup. Broadband white-light illumination from the in-house illumination source of the microscope, OLYMPUS IX 73 (a 12 V–100 W halogen lamp) was used for excitation . The transmitted and scattered signals were collected by a microscope objective and directed to a spectrometer (Andor Kymera 328i). A polarization state generator (PSG), consisting of a rotatable linear polarizer and an achromatic quarter-wave plate (QWP), was used to generate arbitrary polarization states of illumination, while the scattered or transmitted polarization state was analyzed by a polarization state analyzer (PSA) employing a similar QWP–polarizer combination \cite{nayak2025investigatingmmnano2}. All polarization optics were mounted on motorized rotation stages to ensure precise and repeatable measurements.
\par
The full $4 \times 4$ Mueller matrix was reconstructed from 36 polarization projective measurements, providing an overdetermined dataset that improves accuracy and reduces systematic errors, Fig.~\ref{fig1}(a). The reason for choosing this specific methodology is that, alongside the construction of the Mueller matrix, the 36 projective measurements facilitate the observation of the Stokes parameters with different input polarizations. These recorded Stokes parameters provide additional physical insights into the polarized light matter interactions and their manifestation in the specific MM elements \cite{baba2002development36mm}. The Mueller matrix elements are constructed using algebraic manipulations of these polarization projective measurements\cite{baba2002development36mm,nayak2025investigatingmmnano2}. The normalized experimental Mueller matrices ($M_{ij}/M_{11}$) for both enantiomers are presented in Fig.~\ref{fig1}(b), the $M_{11}\equiv 1$ element is not shown. Clear chiro-optical signatures are observed: the CD-related elements ($M_{14}$, $M_{41}$) and CB-related elements ($M_{23}$, $M_{32}$) reverse sign upon switching handedness, confirming an enantiomeric behavior. 
However, finite linear anisotropic contributions are simultaneously present, despite the nominal $C_4$ symmetry of the structure \cite{wang2022pitfalls}. While the circular anisotropic elements invert between enantiomers, the linear anisotropic elements remain nearly unchanged. Furthermore, the ideal symmetry relations expected for a purely chiral, non-depolarizing medium ($M_{14}=M_{41}$ and $M_{23}=-M_{32}$) are only approximately satisfied, indicating a deviation from an ideal chiral medium. This deviation may originate from the presence of linear anisotropies in the system.
\par
\begin{figure*}[htbp]
\centering
    \includegraphics[width=\linewidth]{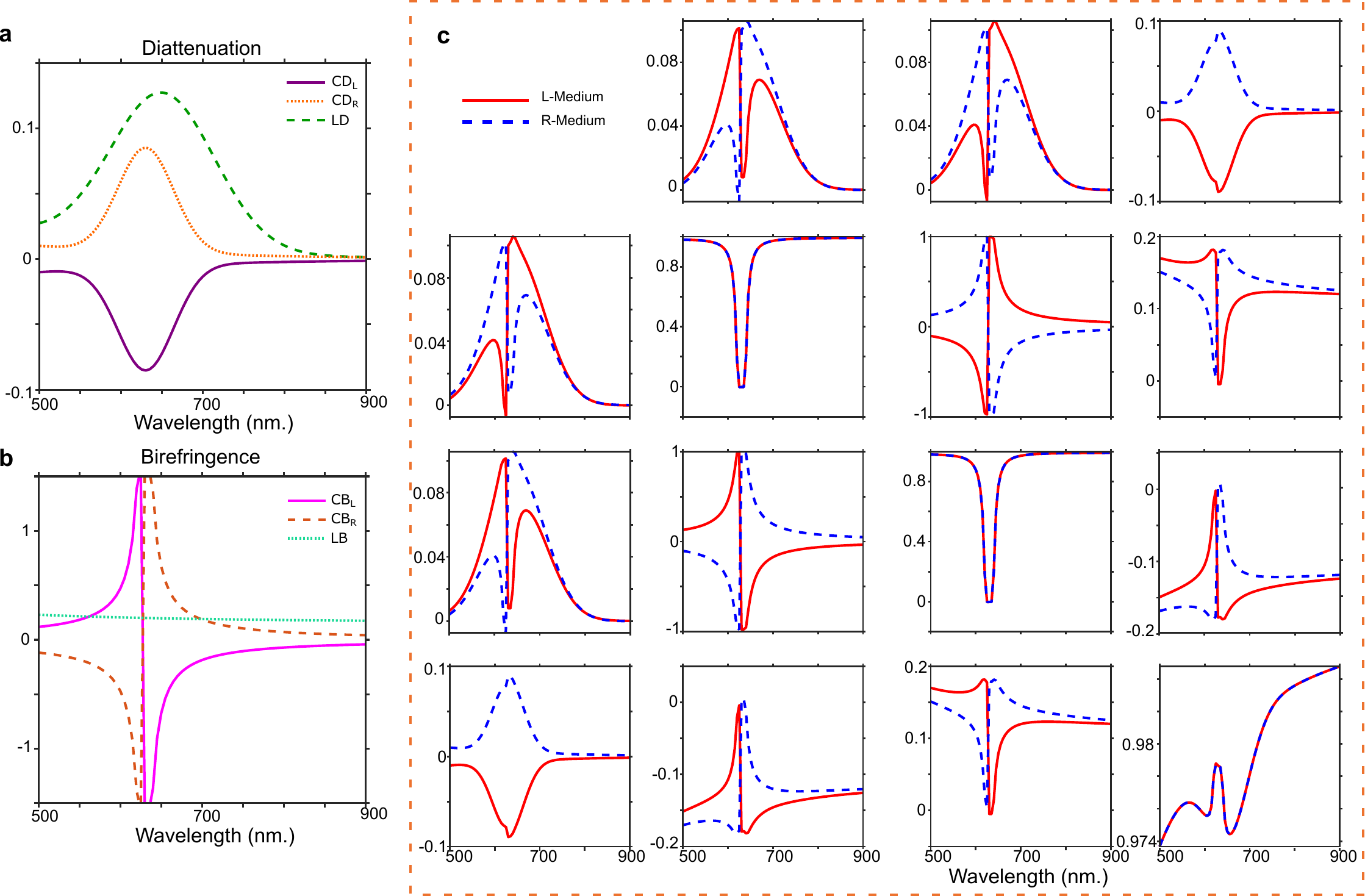}
\caption{Mueller matrix description of a medium exhibiting simultaneous linear and circular anisotropy. 
(a) Spectral variation of the linear diattenuation (LD, green dotted curve) together with circular diattenuation (CD). The CD sign reverses for opposite enantiomers while the LD contribution remains unchanged. 
(b) Spectral variation of the associated linear birefringence (LB) and circular birefringence (CB). 
(c) Resulting Mueller matrices calculated for left- and right-handed media containing both linear and circular anisotropic contributions. Red solid curves denote the left-handed medium and blue dashed curves denote the right-handed medium.}
\label{fig2}
\end{figure*}

To systematically investigate the influence of the concurrent presence of multiple polarization anisotropies in the MM, we model a generalized anisotropic medium incorporating both linear and circular polarization effects. The chiral anisotropic parameters are adopted from the literature \cite{arwin2013cuticle,yao2022extractingchiro}. A broad absorption band with orientation dependence is considered to account for linear dichroism, while a Cauchy-type dispersion is employed to model linear birefringence, as typically observed in biological structures. The linear anisotropic parameters are kept fixed, whereas the sign of the chiral parameters is reversed to represent opposite enantiomers (Fig.~\ref{fig2}(a,b)). 
The resulting Mueller matrices (Fig.~\ref{fig2}(c)) exhibit substantial deviations from those of purely linear or purely circular systems. In particular, the circular diattenuation and circular polarizance are no longer the same $(M_{14} \neq M_{41})$, and the CB-related elements deviate from perfect antisymmetry. Additionally, symmetry relations for linear anisotropic elements (e.g., $M_{12}=M_{21}$, $M_{13}=M_{31}$) are also partially broken. These deviations clearly demonstrate that simultaneous anisotropies generate cross-terms in the Mueller matrix, leading to interdependent signatures across multiple elements. Consequently the standard ellipsometric or chiro-optical measurements fail to probe the intrinsic polarization effects in such systems. For instance in standard approaches, CD is obtained from the differential transmittance under LCP/RCP excitation
$(\frac{LU - RU}{LU + RU})$, which corresponds to the $M_{14}$ element. However in the investigated media the $M_{14}$ element also contains contributions from the coupling between other polarization anisotropies, leading to the inaccurate determination of the intrinsic chirality and other polarization anisotropies of the system, which is discussed next.  

\begin{figure*}[htbp]
\centering
    \includegraphics[width=\linewidth]{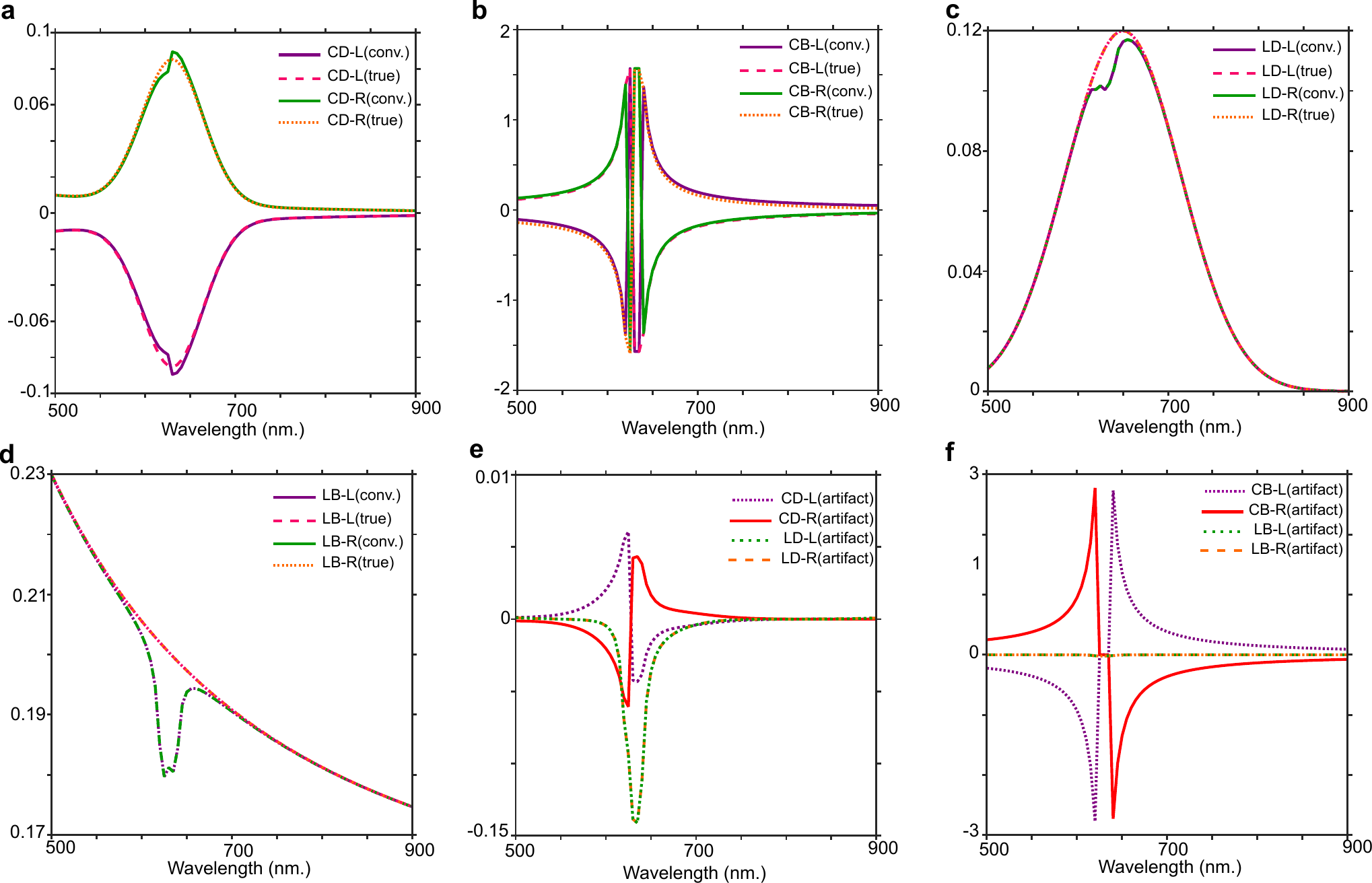}
\caption{Systematic errors in conventional chiro-optical measurements and their influence on chirality retrieval. 
(a–d) Spectral variations of the linear and circular anisotropic parameters obtained using conventional chiro-optical observables (${conv.}$) compared with the intrinsic polarization parameters of the medium (${true}$). The circular diattenuation (CD), circular birefringence (CB), linear diattenuation (LD), and linear birefringence (LB) are shown for both left- (L) and right-handed media. 
(e,f) Artifact plots obtained by subtracting the intrinsic polarization parameters from those retrieved through conventional ellipsometric observables.}
\label{fig3}
\end{figure*}

Figure~\ref{fig3} compares the CD and CB spectra retrieved through conventional ellipsometric observables with the intrinsic parameters of the modeled medium. Significant discrepancies are observed in both magnitude and spectral dispersion. In specific spectral regions, the deviation reaches substantial values. Similar inaccuracies are present for linear anisotropic parameters. The difference between directly interpreted MM elements and intrinsic parameters is shown as artifact plots in Fig.~\ref{fig3}(e,f). The error in the extracted linear diattenuation reaches values of up to $10\%$, while the CB estimation exhibits even more severe artifacts, due to the polarity reversal around the resonant wavelength. Furthermore, when the linear and circular anisotropies possess distinct spectral resonances, the retrieved CD spectrum can exhibit an apparent shift in the peak wavelength relative to its intrinsic position.
These artifacts arise because the measured Mueller matrix elements represent cumulative polarization transformations. When multiple anisotropies coexist, their contributions mix nonlinearly within individual Mueller matrix elements, causing conventional observables to reflect coupled responses rather than isolated intrinsic parameters. Therefore, in systems exhibiting coupled anisotropies, conventional ellipsometric or element-wise MM interpretation can lead to an erroneous quantification of the intrinsic polarization properties.
\par
\begin{figure*}[htbp]
\centering
    \includegraphics[width=\linewidth]{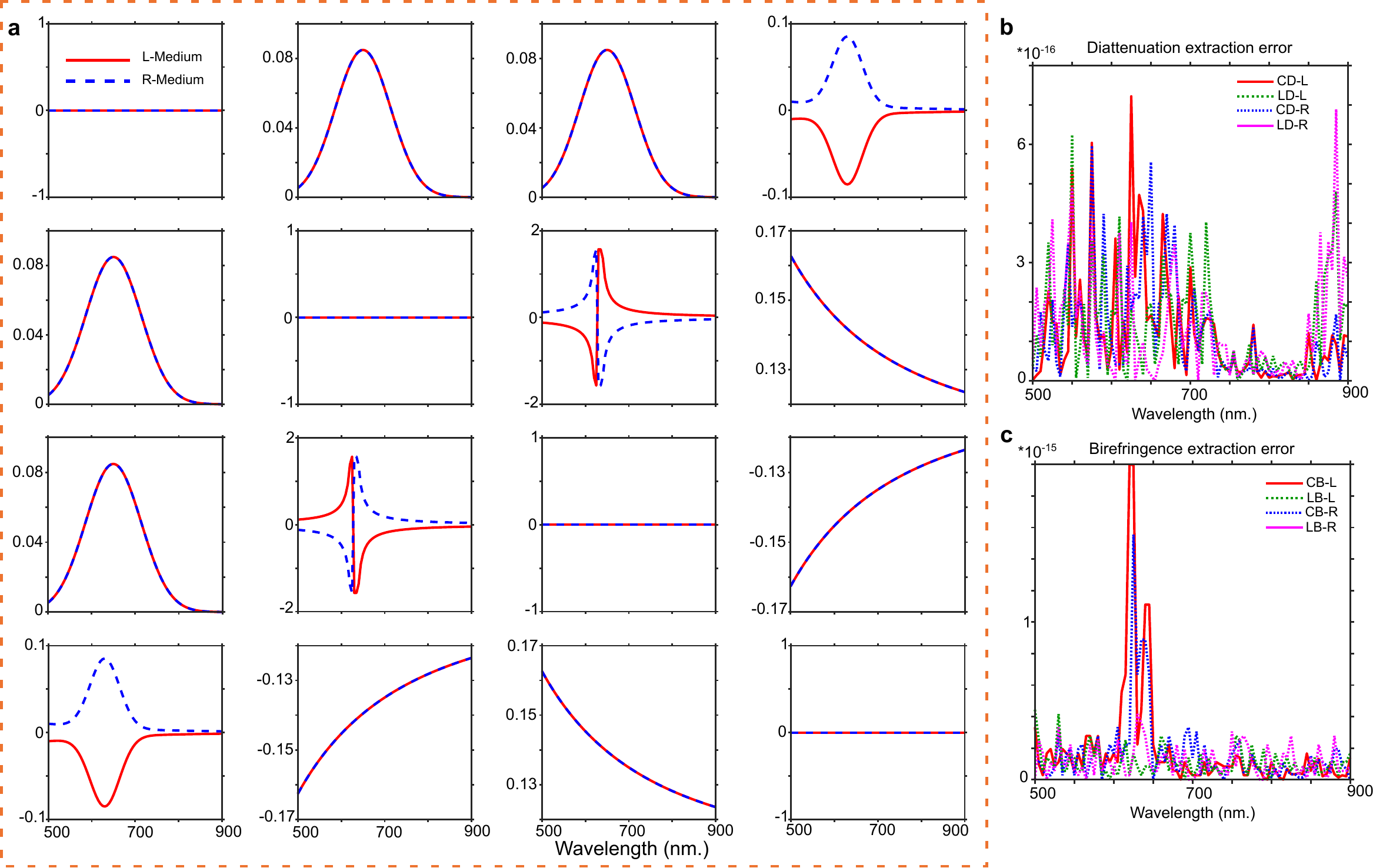}
\caption{Extraction of intrinsic polarization parameters using a differential Mueller matrix decomposition. 
(a) Lorentz antisymmetric matrix ($L_m$) obtained from differential decomposition of the modeled Mueller matrices for left- and right-handed media. Individual polarization effects are decoupled and encoded in distinct matrix elements, exhibiting the expected symmetry relations. 
(b,c) Errors in the parameters retrieved from differential decomposition relative to the intrinsic anisotropies. Panels show deviations for linear and circular diattenuation (b) and birefringence parameters (c).}
\label{fig4}
\end{figure*}

To overcome these limitations, we applied the differential Mueller matrix decomposition to the same synthetic system. The logarithmic matrix $L = \ln M$ was separated into Lorentz antisymmetric ($L_m$) and symmetric ($L_u$) components. Since the synthetic system was assumed to be non-depolarizing, $L_m$ fully characterizes its intrinsic anisotropies. The constructed antisymmetric $L_m$ matrix is shown in \ref{fig4}(a), where all kinds of polarization effects (linear/circular, diattenuation/retardance) present in the system are decoupled and separately manifested in the specific MM elements, as predicted by Eq.~(9).
The extracted polarization parameters exhibit excellent agreement with the given input polarization parameters . The numerical deviation between retrieved and intrinsic parameters is on the order of $10^{-16}$, corresponding to high precision \ref{fig4}(b),(c). This high accuracy arises because the differential decomposition retrieves the generator of the polarization transformation, inherently decoupling simultaneously occurring anisotropies rather than interpreting their cumulative manifestation in individual Mueller matrix elements.
\par
\begin{figure*}[htbp]
\centering
    \includegraphics[width=0.7\linewidth]{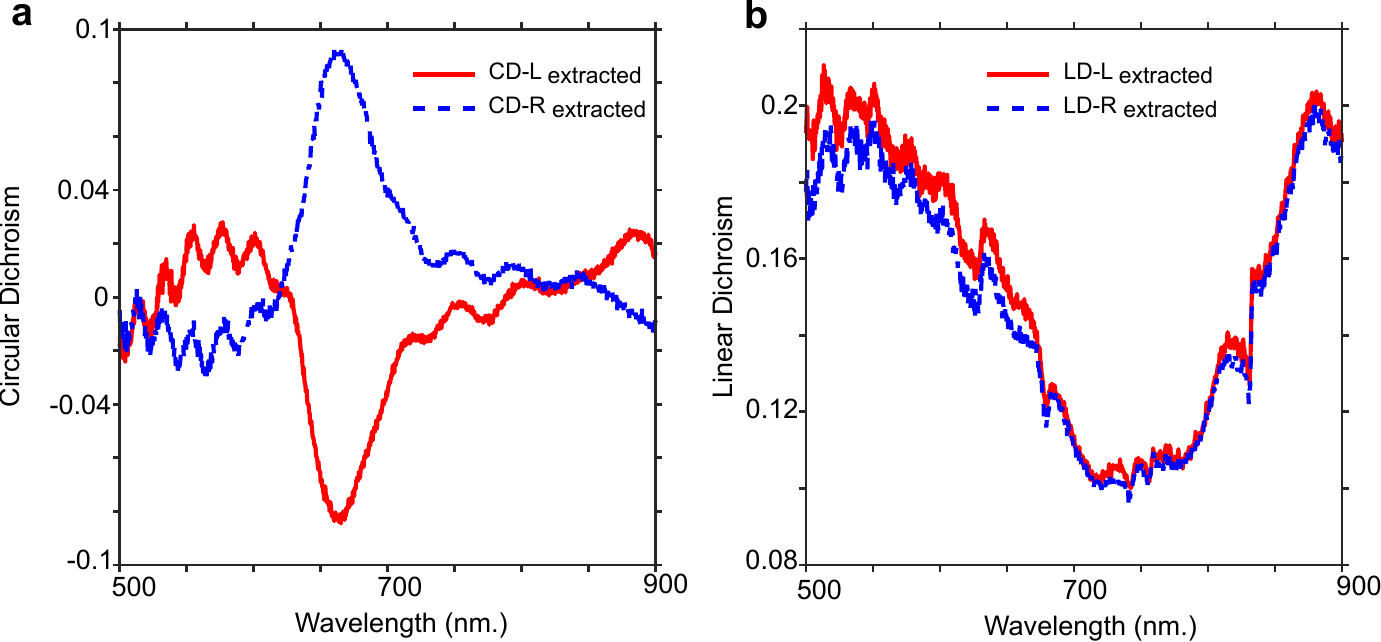}
\caption{Enantiomer discrimination using parameters derived from the differential Mueller matrix decomposition. 
(a) Circular diattenuation spectra extracted from the experimentally measured Mueller matrices of plasmonic gammadion arrays. 
(b) Linear diattenuation obtained from the same decomposition analysis. The circular parameter exhibit mirror symmetry between opposite enantiomers, while the linear contribution remains unchanged.}
\label{fig5}
\end{figure*}

We next applied the differential decomposition to the experimentally measured Mueller matrices of the gammadion arrays. The extracted CD spectra for opposite enantiomers exhibit nearly perfect mirror symmetry (Fig.~\ref{fig5}(a)), restoring the symmetry that was partially obscured in the raw MM elements. On the other hand, linear anisotropic parameters remain invariant under handedness reversal (Fig.~\ref{fig5}(b)).  These results confirm that differential decomposition successfully separates intrinsic circular and linear anisotropies in realistic nanophotonic systems where multiple polarization mechanisms coexist.
\par
\begin{figure*}[htbp]
\centering
    \includegraphics[width=\linewidth]{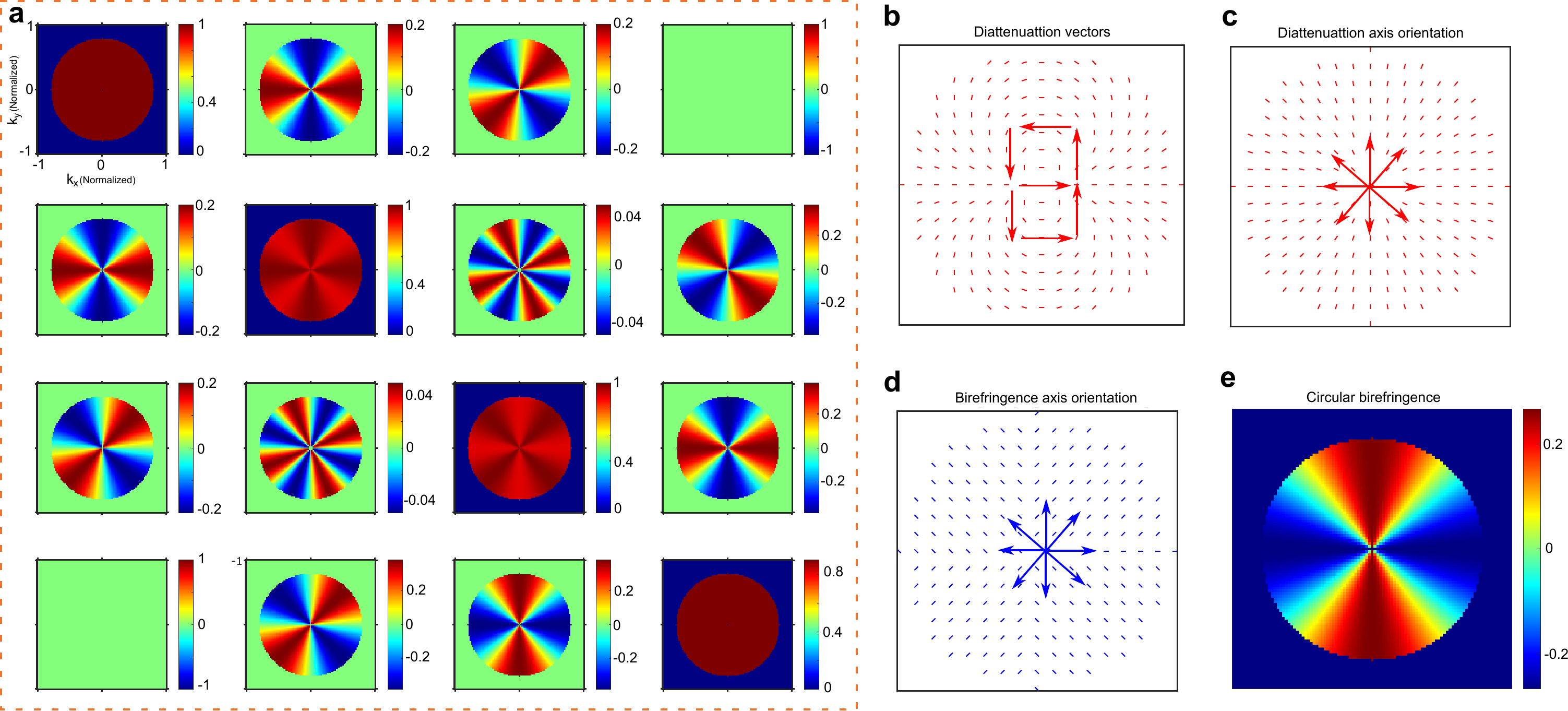}
\caption{Differential decomposition of an inhomogeneous momentum-domain Mueller matrix. 
(a) Mueller matrix describing polarized scattering from a nanostructured medium modeled as a diattenuating retarder with an azimuthally varying orientation axis. 
(b,c) Spatial maps of the diattenuation vector components $(D_x,D_y)$ obtained from decomposition together with the calculated diattenuation magnitude. 
(d) Orientation of the equivalent retarder axis across the momentum plane. 
(e) Wavevector-dependent polarization evolution producing a geometric phase that manifests as an apparent circular birefringence signature, distinct from conventional chiro-optical effects.}
\label{fig6}
\end{figure*}

While the previous analysis focused on spatially homogeneous systems, nanostructured media often exhibit strongly angle-dependent polarization responses. Capturing the polarization-resolved response of such nanostructures—either by varying the angle of incidence or by recording the wavevector-dependent polarized scattering or transmittance—provides additional insight into the underlying light–matter interactions. In this work, we adopt the latter approach, which can be viewed as a momentum-domain polarimetry scheme. To explore the capability of the proposed differential MM framework in such scenarios, we extend the analysis to momentum-resolved Mueller matrix measurements. The Mueller matrix of an inhomogeneous anisotropic scattering system is shown in Fig.~\ref{fig6}(a). The MM elements exhibit azimuthally varying lobe patterns, which are characteristic of polarization-dependent radiation.
The elements $M_{12/21}$ and $M_{13/31}$ exhibit $\cos 2\phi$ and $\sin 2\phi$ angular dependences, indicating the presence of an effective linear diattenuator with azimuthally rotating principal axis. Similar variations in $M_{24/42}$ and $M_{34/43}$ reveal an azimuthally varying linear retarder. Interestingly, the CB-descriptor elements ($M_{23/32}$) show comparable spatial features despite negligible CD elements ($M_{14/41} \approx 0$), suggesting the absence of intrinsic chirality.
The differential decomposition clarifies this behavior: the extracted diattenuation vectors rotate by $4\pi$ over a full azimuthal cycle \ref{fig6}(b), while the corresponding linear diattenuation \ref{fig6}(c), and birefringence \ref{fig6}(d) axes rotate by $2\pi$, reflecting the double-angle relation intrinsic to Stokes–Mueller algebra. This spatially varying anisotropy leads to an opposite geometric phase accumulation for incident LCP and RCP states, a manifestation of spin–orbit interaction of light rather than true circular birefringence.
Importantly, MM polarimetry combined with differential decomposition distinguishes such topology-induced geometric phase effects from intrinsic chiral optical rotation. This demonstrates the capability of the proposed framework not only to quantify polarization anisotropies but also to provide deeper physical insights into complex polarization scattering phenomena.
\section*{Conclusion}
In summary, We have demonstrated that spectroscopic Mueller matrix polarimetry combined with a differential decomposition provides a rigorous framework for extracting intrinsic polarization properties of nanostructured systems exhibiting multiple concurrent anisotropic effects. Using plasmonic gammadion arrays as a model chiral metasurface, we showed that linear and circular anisotropies coexist and couple, producing complex Mueller matrix signatures that cannot be reliably interpreted using conventional ellipsometric or chiro-optical observables.
Through systematic modeling and experiments, we quantified the artifacts that arise when standard approaches are applied to such coupled systems and showed that both the magnitude and spectral dispersion of retrieved parameters can deviate significantly from their intrinsic values. Fortunately, differential decomposition successfully decouples these effects, restoring the expected enantiomeric symmetry of circular parameters while preserving linear anisotropies.
Beyond homogeneous chiral media, we further demonstrated the capability of the framework in momentum-resolved polarimetry, where spatially varying anisotropies and spin–orbit interactions give rise to geometric phase effects that can mimic circular birefringence. Mueller matrix analysis distinctly separated such topology-induced polarization phenomena from intrinsic chiral optical rotation, underscoring the deeper physical insight afforded by full polarimetric characterization. Overall, this work establishes differential Mueller matrix polarimetry as a robust and physically insightful tool for accurate characterization of complex chiral nanophotonic systems, with direct implications for polarization-engineered devices and chiral sensing applications.
\section*{Acknowledgment}
Funding from the Swiss National Science Foundation under project 200021\_212758 is gratefully acknowledged.

\bibliographystyle{unsrtnat}
\bibliography{refs}

@article{li2024opticalpolmani1,
  title={Optical polarization manipulations with anisotropic nanostructures},
  author={Li, Zhancheng and Liu, Wenwei and Zhang, Yuebian and Cheng, Hua and Zhang, Shuang and Chen, Shuqi},
  journal={PhotoniX},
  volume={5},
  number={1},
  pages={30},
  year={2024},
  publisher={Springer}
}

@article{hu2020allpolmani2,
  title={All-dielectric metasurfaces for polarization manipulation: principles and emerging applications},
  author={Hu, Yueqiang and Wang, Xudong and Luo, Xuhao and Ou, Xiangnian and Li, Ling and Chen, Yiqin and Yang, Ping and Wang, Shuai and Duan, Huigao},
  journal={Nanophotonics},
  volume={9},
  number={12},
  pages={3755--3780},
  year={2020},
  publisher={De Gruyter}
}

@article{shen2023reviewpoldevices1,
  title={A review of metasurface polarization devices},
  author={Shen, Zhe and Lin, Xiaojun},
  journal={Optical Materials},
  volume={146},
  pages={114567},
  year={2023},
  publisher={Elsevier}
}

@article{novotny2011antennas,
  title={Antennas for light},
  author={Novotny, Lukas and Van Hulst, Niek},
  journal={Nature photonics},
  volume={5},
  number={2},
  pages={83--90},
  year={2011},
  publisher={Nature Publishing Group UK London}
}

@article{yu2014flatopticspolmetareview,
  title={Flat optics with designer metasurfaces},
  author={Yu, Nanfang and Capasso, Federico},
  journal={Nature materials},
  volume={13},
  number={2},
  pages={139--150},
  year={2014},
  publisher={Nature Publishing Group UK London}
}

@article{shaltout2019spatiotemporalmetareview2,
  title={Spatiotemporal light control with active metasurfaces},
  author={Shaltout, Amr M and Shalaev, Vladimir M and Brongersma, Mark L},
  journal={Science},
  volume={364},
  number={6441},
  pages={eaat3100},
  year={2019},
  publisher={American Association for the Advancement of Science}
}

@article{bliokh2015spinsoi1,
  title={Spin--orbit interactions of light},
  author={Bliokh, Konstantin Yu and Rodr{\'\i}guez-Fortu{\~n}o, Francisco J and Nori, Franco and Zayats, Anatoly V},
  journal={Nature Photonics},
  volume={9},
  number={12},
  pages={796--808},
  year={2015},
  publisher={Nature Publishing Group}
}

@article{valev2013chirality1,
  title={Chirality and chiroptical effects in plasmonic nanostructures: fundamentals, recent progress, and outlook},
  author={Valev, Ventsislav K and Baumberg, Jeremy J and Sibilia, Concita and Verbiest, Thierry},
  journal={Advanced Materials},
  volume={25},
  number={18},
  pages={2517--2534},
  year={2013},
  publisher={Wiley Online Library}
}

@article{electromagneticchiralityreview1,
  title={Electromagnetic chirality: from fundamentals to nontraditional chiroptical phenomena},
  author={Mun, Jungho and Kim, Minkyung and Yang, Younghwan and Badloe, Trevon and Ni, Jincheng and Chen, Yang and Qiu, Cheng-Wei and Rho, Junsuk},
  journal={Light: Science \& Applications},
  volume={9},
  number={1},
  pages={139},
  year={2020},
  publisher={Nature Publishing Group UK London}
}

@article{liu2023detectionbiomreview3sensing,
  title={Detection and analysis of chiral molecules as disease biomarkers},
  author={Liu, Yaoran and Wu, Zilong and Armstrong, Daniel W and Wolosker, Herman and Zheng, Yuebing},
  journal={Nature Reviews Chemistry},
  volume={7},
  number={5},
  pages={355--373},
  year={2023},
  publisher={Nature Publishing Group UK London}
}

@article{im2024perspectivessensingreview1,
  title={Perspectives of chiral nanophotonics: from mechanisms to biomedical applications},
  author={Im, Seongmin and Mousavi, Seyedehniousha and Chen, Yun-Sheng and Zhao, Yang},
  journal={npj Nanophotonics},
  volume={1},
  number={1},
  pages={46},
  year={2024},
  publisher={Nature Publishing Group UK London}
}

@article{peng2016chirallasing,
  title={Chiral modes and directional lasing at exceptional points},
  author={Peng, Bo and {\"O}zdemir, {\c{S}}ahin Kaya and Liertzer, Matthias and Chen, Weijian and Kramer, Johannes and Y{\i}lmaz, Huzeyfe and Wiersig, Jan and Rotter, Stefan and Yang, Lan},
  journal={Proceedings of the National Academy of Sciences},
  volume={113},
  number={25},
  pages={6845--6850},
  year={2016},
  publisher={National Academy of Sciences}
}

@article{chvojka2020visiblepolarizationmultiplexing,
  title={Visible light communications: increasing data rates with polarization division multiplexing},
  author={Chvojka, Petr and Burton, Andrew and Pesek, Petr and Li, Xicong and Ghassemlooy, Zabih and Zvanovec, Stanislav and Anthony Haigh, Paul},
  journal={Optics Letters},
  volume={45},
  number={11},
  pages={2977--2980},
  year={2020},
  publisher={Optical Society of America}
}

@article{martinez2018polarimetrynano,
  title={Polarimetry enabled by nanophotonics},
  author={Martinez, Alejandro},
  journal={Science},
  volume={362},
  number={6416},
  pages={750--751},
  year={2018},
  publisher={American Association for the Advancement of Science}
}

@article{lamon2024neuromorphicdata,
  title={Neuromorphic optical data storage enabled by nanophotonics: a perspective},
  author={Lamon, Simone and Zhang, Qiming and Yu, Haoyi and Gu, Min},
  journal={ACS Photonics},
  volume={11},
  number={3},
  pages={874--891},
  year={2024},
  publisher={ACS Publications}
}

@article{li2024opticalneedpol,
  title={Optical polarization manipulations with anisotropic nanostructures},
  author={Li, Zhancheng and Liu, Wenwei and Zhang, Yuebian and Cheng, Hua and Zhang, Shuang and Chen, Shuqi},
  journal={PhotoniX},
  volume={5},
  number={1},
  pages={30},
  year={2024},
  publisher={Springer}
}

@article{shao2022analyzingneedpol2,
  title={Analyzing the influence of imaging resolution on polarization properties of scattering media obtained from Mueller matrix},
  author={Shao, Conghui and Chen, Binguo and He, Honghui and He, Chao and Shen, Yuanxing and Zhai, Haoyu and Ma, Hui},
  journal={Frontiers in Chemistry},
  volume={10},
  pages={936255},
  year={2022},
  publisher={Frontiers Media SA}
}

@article{doeleman2018experimentalpolbic,
  title={Experimental observation of a polarization vortex at an optical bound state in the continuum},
  author={Doeleman, Hugo M and Monticone, Francesco and den Hollander, Wouter and Al{\`u}, Andrea and Koenderink, A Femius},
  journal={Nature Photonics},
  volume={12},
  number={7},
  pages={397--401},
  year={2018},
  publisher={Nature Publishing Group UK London}
}

@article{he2022towardsstructuredlight,
  title={Towards higher-dimensional structured light},
  author={He, Chao and Shen, Yijie and Forbes, Andrew},
  journal={Light: Science \& Applications},
  volume={11},
  number={1},
  pages={205},
  year={2022},
  publisher={Nature Publishing Group UK London}
}

@article{ray2017polarizationmmnano1,
  title={Polarization-tailored Fano interference in plasmonic crystals: a Mueller matrix model of anisotropic Fano resonance},
  author={Ray, Subir K and Chandel, Shubham and Singh, Ankit K and Kumar, Abhishek and Mandal, Arpita and Misra, Subhradeep and Mitra, Partha and Ghosh, Nirmalya},
  journal={ACS nano},
  volume={11},
  number={2},
  pages={1641--1648},
  year={2017},
  publisher={ACS Publications}
}

@article{kawabata2014fundamentalsellipsometry,
  title={Fundamentals and applications of ellipsometry},
  author={Kawabata, S},
  journal={Hyomen Kagaku},
  volume={35},
  pages={286--293},
  year={2014}
}

@article{losurdo2009spectroscopicellipsometry2,
  title={Spectroscopic ellipsometry and polarimetry for materials and systems analysis at the nanometer scale: state-of-the-art, potential, and perspectives},
  author={Losurdo, Maria and Bergmair, Michael and Bruno, Giovanni and Cattelan, Denis and Cobet, Christoph and De Martino, Antonello and Fleischer, Karsten and Dohcevic-Mitrovic, Zorana and Esser, Norbert and Galliet, Melanie and others},
  journal={Journal of Nanoparticle Research},
  volume={11},
  number={7},
  pages={1521--1554},
  year={2009},
  publisher={Springer}
}

@article{hentschel2017chiralchiro-opticalnano,
  title={Chiral plasmonics},
  author={Hentschel, Mario and Sch{\"a}ferling, Martin and Duan, Xiaoyang and Giessen, Harald and Liu, Na},
  journal={Science advances},
  volume={3},
  number={5},
  pages={e1602735},
  year={2017},
  publisher={American Association for the Advancement of Science}
}

@article{yao2022extractingchiro,
  title={Extracting pure circular dichroism from hierarchically structured CdS magic cluster films},
  author={Yao, Yuan and Ugras, Thomas J and Meyer, Talisi and Dykes, Matthew and Wang, Da and Arbe, Arantxa and Bals, Sara and Kahr, Bart and Robinson, Richard D},
  journal={ACS nano},
  volume={16},
  number={12},
  pages={20457--20469},
  year={2022},
  publisher={ACS Publications}
}

@article{abasahl2021fabrication,
  title={Fabrication of plasmonic structures with well-controlled nanometric features: a comparison between lift-off and ion beam etching},
  author={Abasahl, Banafsheh and Santschi, Christian and Raziman, TV and Martin, Olivier JF},
  journal={Nanotechnology},
  volume={32},
  number={47},
  pages={475202},
  year={2021},
  publisher={IOP Publishing}
}

@article{chen2021principles,
  title={Principles, functions, and applications of optical meta-lens},
  author={Chen, Mu Ku and Wu, Yongfeng and Feng, Lei and Fan, Qingbin and Lu, Minghui and Xu, Ting and Tsai, Din Ping},
  journal={Advanced Optical Materials},
  volume={9},
  number={4},
  pages={2001414},
  year={2021},
  publisher={Wiley Online Library}
}

@article{bandaru2022microstructure,
  title={Microstructure dictates the behavior of plasmons in nanostructured Ag-Cu alloy films},
  author={Bandaru, Pravallika and Ummethala, Govind and Malladi, Sai Rama Krishna and Dutta-Gupta, Shourya},
  journal={The Journal of Physical Chemistry C},
  volume={126},
  number={37},
  pages={15915--15923},
  year={2022},
  publisher={ACS Publications}
}

@book{goldstein2017polarized,
  title={Polarized light},
  author={Goldstein, Dennis H},
  year={2017},
  publisher={CRC press}
}

@article{parrish2025differentialldlbcoupling,
  title={Differential absorption of circularly polarized light by a centrosymmetric crystal},
  author={Parrish, Katherine A and Salij, Andrew and Kamp, Kendall R and Smith, Evan and Utama, M Iqbal Bakti and Bergsten, Anders J and Czerwinski, Rachel and Smith, Mackinsey A and Hersam, Mark C and Poeppelmeier, Kenneth R and others},
  journal={Science},
  volume={388},
  number={6752},
  pages={1194--1197},
  year={2025},
  publisher={American Association for the Advancement of Science}
}

@article{yu2014flatmultipolnano,
  title={Flat optics with designer metasurfaces},
  author={Yu, Nanfang and Capasso, Federico},
  journal={Nature materials},
  volume={13},
  number={2},
  pages={139--150},
  year={2014},
  publisher={Nature Publishing Group UK London}
}

@article{schaferling2012tailoringchiro-opticalnano2,
  title={Tailoring enhanced optical chirality: design principles for chiral plasmonic nanostructures},
  author={Sch{\"a}ferling, Martin and Dregely, Daniel and Hentschel, Mario and Giessen, Harald},
  journal={Physical Review X},
  volume={2},
  number={3},
  pages={031010},
  year={2012},
  publisher={APS}
}

@book{schubert2004infraredellipsometry3,
  title={Infrared ellipsometry on semiconductor layer structures: phonons, plasmons, and polaritons},
  author={Schubert, Mathias},
  volume={209},
  year={2004},
  publisher={Springer Science \& Business Media}
}

@article{nayak2025investigatingmmnano2,
  title={Investigating spin-orbit interaction of light in micro-and nanophotonic systems using polarization Mueller matrix},
  author={Nayak, Jeeban Kumar and Guchhait, Shyamal and Banerjee, Ayan and Gupta, Subhasish Dutta and Ghosh, Nirmalya},
  journal={Applied Physics Letters},
  volume={126},
  number={3},
  year={2025},
  publisher={AIP Publishing}
}

@book{gil2022polarizedpolbook1,
  title={Polarized light and the Mueller matrix approach},
  author={Gil, Jos{\'e} Jorge and Ossikovski, Razvigor and Gil, Jose J},
  year={2022},
  publisher={CRC press}
}

@article{arwin2013cuticle,
  title={Cuticle structure of the scarab beetle Cetonia aurata analyzed by regression analysis of Mueller-matrix ellipsometric data},
  author={Arwin, Hans and Berlind, Torun and Johs, Blaine and J{\"a}rrendahl, Kenneth},
  journal={Optics Express},
  volume={21},
  number={19},
  pages={22645--22656},
  year={2013},
  publisher={Optical Society of America}
}

@article{baba2002development36mm,
  title={Development and calibration of an automated Mueller matrix polarization imaging system},
  author={Baba, Justin S and Chung, Jung-Rae and DeLaughter, Aimee H and Cameron, Brent D and Cote, Gerard L},
  journal={Journal of biomedical optics},
  volume={7},
  number={3},
  pages={341--349},
  year={2002},
  publisher={Society of Photo-Optical Instrumentation Engineers}
}

@article{arteaga2009pseudopolar,
  title={Pseudopolar decomposition of the Jones and Mueller--Jones exponential polarization matrices},
  author={Arteaga, Oriol and Canillas, Adolf},
  journal={Journal of the Optical Society of America A},
  volume={26},
  number={4},
  pages={783--793},
  year={2009},
  publisher={Optical Society of America}
}

@article{lu1996interpretationpolardecomp1,
  title={Interpretation of Mueller matrices based on polar decomposition},
  author={Lu, Shih-Yau and Chipman, Russell A},
  journal={Journal of the Optical Society of America A},
  volume={13},
  number={5},
  pages={1106--1113},
  year={1996},
  publisher={Optical Society of America}
}

@article{azzam1978propagationdiffdecomp1,
  title={Propagation of partially polarized light through anisotropic media with or without depolarization: a differential 4$\times$ 4 matrix calculus},
  author={Azzam, RMA},
  journal={Journal of the Optical Society of America},
  volume={68},
  number={12},
  pages={1756--1767},
  year={1978},
  publisher={Optical Society of America}
}

@article{ossikovski2011differential,
  title={Differential matrix formalism for depolarizing anisotropic media},
  author={Ossikovski, Razvigor},
  journal={Optics letters},
  volume={36},
  number={12},
  pages={2330--2332},
  year={2011},
  publisher={Optical Society of America}
}

@article{miao2025opticalchiro-opticalmm,
  title={Optical parameter determination in chiral media: Refractive index and Pasteur parameter insights},
  author={Miao, Qiwei and Jia, Fengyang and Chen, Jiajie and Wang, Xin and Sun, Lixun and Lu, Fanfan and Zhang, Wending and Mei, Ting},
  journal={Optics \& Laser Technology},
  volume={192},
  pages={114028},
  year={2025},
  publisher={Elsevier}
}

@article{lightner2023understandingchiralartifact1,
  title={Understanding artifacts in chiroptical spectroscopy},
  author={Lightner, Carin R and Desmet, Filip and Gisler, Daniel and Meyer, Stefan A and Perez Mellor, Ariel Francis and Niese, Hannah and Rosspeintner, Arnulf and Keitel, Robert C and Burgi, Thomas and Herrebout, Wouter A and others},
  journal={ACS Photonics},
  volume={10},
  number={2},
  pages={475--483},
  year={2023},
  publisher={ACS Publications}
}

@article{ugras2023canchiralartifat2,
  title={Can we still measure circular dichroism with circular dichroism spectrometers: The dangers of anisotropic artifacts},
  author={Ugras, Thomas J and Yao, Yuan and Robinson, Richard D},
  journal={Chirality},
  volume={35},
  number={11},
  pages={846--855},
  year={2023},
  publisher={Wiley Online Library}
}

@book{gupta2015waveoptics,
  title={Wave optics: Basic concepts and contemporary trends},
  author={Gupta, Subhasish Dutta and Ghosh, Nirmalya and Banerjee, Ayan},
  year={2015},
  publisher={CRC Press}
}

@article{kumar2012comparativepolarvsdiff,
  title={Comparative study of differential matrix and extended polar decomposition formalisms for polarimetric characterization of complex tissue-like turbid media},
  author={Kumar, Satish and Purwar, Harsh and Ossikovski, Razvigor and Vitkin, I Alex and Ghosh, Nirmalya},
  journal={Journal of biomedical optics},
  volume={17},
  number={10},
  pages={105006--105006},
  year={2012},
  publisher={Society of Photo-Optical Instrumentation Engineers}
}

@article{kumar2024muellerpolarvsdiff,
  title={Mueller matrix-based characterization of cervical tissue sections: a quantitative comparison of polar and differential decomposition methods},
  author={Kumar, Nishkarsh and Kumar Nayak, Jeeban and Pradhan, Asima and Ghosh, Nirmalya},
  journal={Journal of Biomedical Optics},
  volume={29},
  number={5},
  pages={052916--052916},
  year={2024},
  publisher={Society of Photo-Optical Instrumentation Engineers}
}

@article{wang2022pitfalls,
  title={Pitfalls in the spectral measurements of polarization-altering metasurfaces},
  author={Wang, Hsiang-Chu and Martin, Olivier JF},
  journal={Applied Optics},
  volume={61},
  number={27},
  pages={8100--8107},
  year={2022},
  publisher={Optica Publishing Group}
}

\end{document}